\documentclass[aps,prl,twocolumn,showpacs,amsmath,amssymb]{revtex4}
\usepackage{graphics}
\newcommand{\etal}{\textit{et al.}}
\begin{document}

\title{Spectral-fluctuations test of the quark-model baryon spectrum}

\author{C. \surname{Fern\'andez-Ram\'{\i}rez}}
\altaffiliation{Corresponding author: Baryons}
\email{cesar@nuc2.fis.ucm.es}
\affiliation{Instituto de Estructura de la Materia, 
CSIC, Serrano 123, E-28006 Madrid, Spain}

\author{A. \surname{Rela\~no}}
\altaffiliation{Corresponding author: Quantum chaos and statistical tools}
\email{armando@iem.cfmac.csic.es}
\affiliation{Instituto de Estructura de la Materia, 
CSIC, Serrano 123, E-28006 Madrid, Spain}

\date{\today}
\begin{abstract}
We study the low-lying baryon spectrum (up to 2.2 GeV) provided by
experiments and different quark models using statistical tools which
allow to postulate the existence of missing levels in spectra. We
confirm that the experimental spectrum is compatible with Random
Matrix Theory --the paradigmatic model of quantum chaos--, and we find
that the quark models are more similar to a Poisson distribution,
which is not compatible with what should be expected in a correlated
spectrum.  From our analysis it stems that the spectral fluctuation
properties of quark-model spectra are incompatible with experimental
data.  This result can be used to enlighten the problem of
\textit{missing} resonances.
\end{abstract}
\maketitle

%%%%%%%%%%%%%%%%%
% Introduccion %
%%%%%%%%%%%%%%%%%

Since the discovery of the first excited state of the nucleon, 
the $\Delta$(1232) \cite{Anderson52}, 
baryons have played a central role in the study of the strong interaction.
After this discovery, baryons proliferated and
their mass spectrum attracted a lot of attention. 
In the 60's and 70's, nonrelativistic quark models
were worked out \cite{Quark} and
the quark model framework was established, 
contributing strongly to
the development of Quantum Chromodynamics (QCD).
Later on, quark models evolved into relativistic
versions \cite{Capstick86,Capstick00,Bonn01}.

It is well known that the number of baryons predicted by quark models
\cite{Quark,Capstick86,Capstick00,Bonn01} is substantially larger than
what is observed in meson scattering and production experiments
\cite{PDG2006}.  This fact raises the problem of \textit{missing}
resonances, which has opened the door to a huge experimental effort in
the last years to observe and identify these missing states
\cite{experiments}.  These experiments have to achieve high precision
due to the important background (which can veil resonances)
and the overlap of baryons, as well as the need to survey different
meson production channels and observables.  The procedure to assess
the existence of these elusive baryons consists on analyses of partial waves
\cite{SAID} and polarization observables \cite{dutta} of the
reactions comparing experimental data from different sources to what
is obtained after including or removing the hypothetical resonance.  If
data are better reproduced, the existence of a resonance is possible
but sometimes debatable (the pentaquark search is a very good example
of how difficult these studies are \cite{pentaquark}).  The Particle Data
Group (PDG) rates the possible existence of the resonances based on
the quantity and quality of experimental data.  Only after several
independent experiments and analyses, a baryon is awarded a
\textit{well-established} status, rated with three or four stars.

In the last years, Lattice QCD (LQCD) \cite{lattice}
and effective QCD-inspired models (EQCDiM) \cite{felipe}
have started to provide insight on
the non-perturbative energy regime of QCD
where the low-lying baryons live, 
but they are still far away from providing a
complete analysis.
Hence, we have to resort to tractable models, such as 
quark models, to study the baryon spectrum.
For this reason, quark models are expected
to retain in forthcoming years the importance they had 
in the study of the static
properties and internal structure of baryons.

%%%%%%%%%%%%%%%%%%%%%%%%%%%%%%%%%%%%%%%%%  
% Proposito de la Letter. Antecedentes. %
%%%%%%%%%%%%%%%%%%%%%%%%%%%%%%%%%%%%%%%%%

It is well established that baryons are aggregates of partons. 
Consequently, the mass spectrum of low-lying baryons is an energy
spectrum of a standard many-body quantum system, like an atomic nucleus,
and it consists on all the possible bound and excited
states which stem from an interacting  many-body quantum system.
Since Wigner discovered that the statistical
properties of complex nuclear spectra are well described by Random
Matrix Theory (RMT) \cite{Porter}, statistical methods have become a
powerful tool to study the energy spectra of quantum systems. The most
striking result in this field is that the statistical properties of
the energy-level fluctuations are universal and determine if a
system is chaotic or integrable. Integrable systems display a
non-correlated sequence of levels, which follows the Poisson distribution
\cite{Berry77}, whereas chaotic systems are characterized by a
correlation structure described by RMT \cite{Bohigas84}. This kind of
analysis has been already applied to the hadron mass 
spectrum by Pascalutsa \cite{Pascalutsa03}
obtaining a chaotic-like behavior.

In this Letter we apply the spectral statistical techniques
to the problem of missing resonances. Recent works by
Bohigas and Pato \cite{Bohigas04} prove that if 
we randomly remove some energy levels from a correlated spectrum,
it partially loses correlations and becomes closer to a Poisson
distribution
regardless its actual correlation structure. 
Following this result, we
can obtain hints on the existence of missing resonances comparing
the spectral correlations of the baryon spectra 
supplied by experiments and quark models. 
We have analyzed the experimental spectrum (Breit-Wigner masses)
provided by the PDG 
(set EXP in what follows) \cite{PDG2006}
and the spectra given by the relativistic quark models by
Capstick and Isgur (set CI) \cite{Capstick86} and by L\"oring 
\etal~\cite{Bonn01} (sets L1 and L2 which correspond, respectively,
to models $\cal{A}$ and $\cal{B}$ in \cite{Bonn01}).
We have considered all the resonance states in these spectra up to 2.2 GeV.
In principle, if the experimental spectrum is not complete due to 
the absence of some baryons which are not observed but
are predicted by quark models, and if we assume that missing resonances
are randomly distributed \cite{nota1}, the experimental spectrum
should be less correlated than the theoretical ones.

%%%%%%%%%%%%%%%%%%%%%%%%%%%%%%%%%%%%%%%
% Analisis preliminar. Procedimiento. %
%%%%%%%%%%%%%%%%%%%%%%%%%%%%%%%%%%%%%%%

Prior to any statistical analysis we have to accomplish two
preliminary tasks.  First of all, it is necessary to identify the
different symmetries involved in the spectra. If a sequence of levels
involves more than one symmetry, its spectral statistics are deflected
towards a Poisson distribution (see \cite{Porter,mixing_symmetries}
for generic reviews and \cite{Molina06} for a recent work where the
effects of both mixing symmetries and missing levels in the same
sequence are surveyed).  Hence, it is necessary to extract from the
full spectrum sequences of levels involving the same symmetries
(quantum numbers) to proceed with the spectral analysis.

In the spectra considered in this Letter, we can identify the
following symmetries associated to the baryons: 
spin, isospin, parity, and strangeness. 
Strangeness can be dropped due to the assumption 
of flavor $SU(3)$ invariance. Therefore, for every statistical analysis,
we split all the spectra in sequences where all the levels
present the same values of spin, isospin, and parity.
For reasons stated below we only account
 for sequences with three or more levels.

The second preliminary task is the \textit{unfolding} procedure. 
In any energy-level spectrum, 
we can split the level density, $\rho(E)$,
into a smooth part, $\overline{\rho}(E)$,
and a fluctuating part, $\widetilde{\rho}(E)$, with
 $\rho(E) = \overline{\rho}(E) + \widetilde{\rho}(E)$.
The unfolding procedure allows to extract the fluctuating part
from the level density, removing the smooth component of the spectrum.
There are several ways to unfold a spectrum and we choose
the simplest one. First, we compute the
distance between two consecutive levels, $S_i = E_{i+1} - E_i$, and
then we rescale $S_i$ using its average value $s_i =S_i / \left<S
\right>$ \cite{nota3}.
The resulting quantities are called Nearest Neighbor
Spacings (NNS). This procedure undergoes some problems, specially
in the long-range correlation analysis \cite{Gomez02}, but it
is suitable for the kind of analysis of our concern in this Letter.

%%%%%%%%%%%%%%%%%%%%%%%%%%%%%%%%%%%%
% Analisis preliminar. Resultados. %
%%%%%%%%%%%%%%%%%%%%%%%%%%%%%%%%%%%%

From the NNS we obtain
one of the most relevant quantities in spectral statistical analysis:
the Nearest Neighbor Spacing Distribution (NNSD). 
The NNSD follows the Poisson distribution $P(s) = \exp(-s)$ 
if the spectrum is integrable (non-correlated) \cite{Berry77}, 
but it follows the Wigner surmise
$P(s)=\frac{\pi s}{2} \exp \left(-\frac{\pi s^2}{4} \right)$,
which stems from RMT,
for a chaotic (correlated) spectrum \cite{Bohigas84}.
For our purpose here, it is enough to consider that the less (more) 
correlated the sequence of levels is, the closer to the 
Poisson (Wigner) distribution the NNSD is.

In order to obtain a significative result we have calculated the NNSD,
$P(s)$, for each one of the four different spectra we survey: 
EXP, CI, L1, and L2.
We account for all the sequences $\{ s_i \}_X$, where $X$
stands for the quantum numbers which identify each sequence \cite{nota2}.
Set EXP has 70 energy levels distributed in 15 sequences;
set CI, 145 levels and 19 sequences;
set L1, 142 levels and 21 sequences;
and set L2, 104 levels and 19 sequences. 

We also have evaluated the function
\begin{equation}
F(x)=1-\int_0^x ds \; P(s),
\end{equation}
which is related to the accumulated NNSD
and allows a better study of the tail of the distribution.

\begin{figure}
\begin{center}
\rotatebox{-90}{\scalebox{0.2}[0.25]{\includegraphics{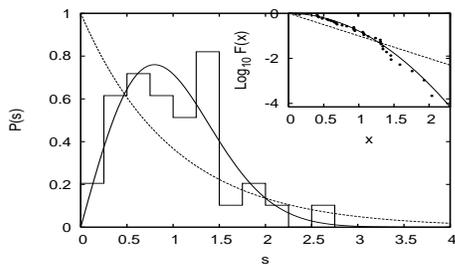}}} \\
(a) Experimental values from PDG (set EXP). \\
\rotatebox{-90}{\scalebox{0.2}[0.25]{\includegraphics{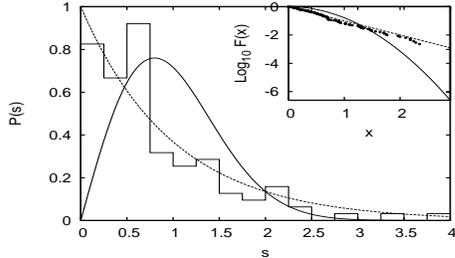}}} \\ 
(b) Model by Capstick and Isgur (set CI). \\
\rotatebox{-90}{\scalebox{0.2}[0.25]{\includegraphics{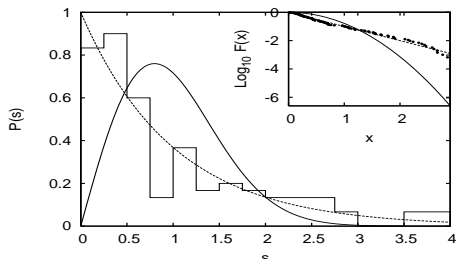}}} \\
(c) Model by L\"oring \etal~ (set L1). \\
\rotatebox{-90}{\scalebox{0.2}[0.25]{\includegraphics{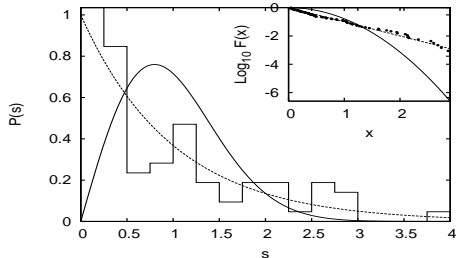}}} \\
(d) Model by L\"oring \etal~ (set L2). \\
\caption[]{NNSD for the experimental spectrum provided by the PDG data, 
the model by Capstick and Isgur, and two parametrizations of the model 
by L\"oring \etal~ The histogram represents the spacings;
the solid line, the Wigner surmise; and the dashed line, the Poisson
distribution. The inset shows the function $F(x)$ in logarithmic
scale.} 
\label{fig:pds_exp}
\end{center}
\end{figure}

Results are depicted in Fig. \ref{fig:pds_exp}. The experimental
spectrum is closer to the Wigner surmise 
than to the Poisson distribution
(in good agreement with what Pascalutsa obtains in \cite{Pascalutsa03}).
On the other hand, the three theoretical spectra are
quite close to the Poisson distribution, so that they are
less correlated than the experimental spectrum. These results are
opposite to what is expected from an experimental spectrum spoiled
by missing resonances. If the experimental spectrum is not complete
due to missing states, it has to be closer to the Poisson
distribution than the theoretical ones.

However, there is one subtle objection to our analysis
regarding the unfolding procedure which has to be considered in detail.  
As we have
pointed before, the unfolding has to be performed independently for
each $\{ s_i \}_X$ sequence. This entails that, at least in some
cases, we have worked with very short sequences of levels.
In such cases, the
unfolding procedure could give rise to misleading results if it makes the
spacings $s_i$ spuriously closer to $s_i \simeq 1$, shifting the NNSD
far away from the Poisson distribution and bringing it to the
Wigner surmise (roughly centered at $s = 1$). As an extreme
example, let us see what happens to a sequence of two levels after
the unfolding. In this case, we only have 
$S_1 = E_2 - E_1$, and, therefore, $s_1 = S_1 / \left< S \right> =1$. 
This is the reason why we have included only sequences
with three or more levels in our analysis.
Moreover, since the experimental sequences are shorter than
the theoretical ones, they can be more affected by this problem, spoiling
the direct comparison we have made above. Therefore, 
to reach a final conclusion it is mandatory to perform an improved
analysis.

To elucidate whether the unfolding procedure can give rise to misleading
results, we have accomplished a more sophisticated analysis of the
NNSD. Since the unfolding could yield different
effects on experimental and theoretical spectra, we have avoided a
direct comparison among them. Instead, we have built 
four Poisson-like and four RMT-like reference spectra, each of them
optimized to study each one of the four sets of data (EXP, CI, L1, and L2).
We have proceeded in the following way for every set.
For each sequence $\{s_i \}_X$ of a given spectrum we have built
two equivalent reference sequences $\{s_i' \}_X$ with the same length,
one RMT-like and another Poisson-like.
In doing so, we start from one long 
RMT-like spectrum and one long Poisson-like spectrum.
We divide each spectrum in as many subsequences as sets with the same quantum
numbers the spectrum under consideration has, each one with its
appropiate length.
Next, we unfold each reference sequence independently and
calculate the NNSD for each set.
Finally, we compare each spectrum obtained
directly from the experimental data (EXP) and the quark models
(CI, L1, and L2) to the two \textit{ad hoc} built-up RMT-like and Poisson-like
reference spectra.
In this way, each reference spectra is distorded 
by the unfolding in the same way as
sets EXP, CI, L1, and L2 are.
Consequently, we can verify whether their correlations are true
or spurious due to the unfolding. 

\begin{table}
\begin{ruledtabular}
\caption[]{Probability to obtain, under the null hypothesis, a
value of the Kolmogorov-Smirnov test statistic as extreme as that
observed.}
\begin{tabular}{c||c|c|c|c}
Spectrum & EXP & CI & L1 & L2 \\ \hline
Poisson & $0.25$ & $0.44$ & $0.37$ & $0.10$ \\
RMT & $0.56$ & $4.5 \cdot 10^{-5}$ & $1.3 \cdot 10^{-4}$ & $6.5 \cdot 10^{-3}$
\end{tabular}
\label{tab:KS}
\end{ruledtabular}
\end{table}

In order to obtain significative results, we have performed the
Kolmogorov-Smirnov distribution test \cite{NAG} in each case. 
As null hypothesis we have chosen
that the studied NNSD coincides with the \textit{ad hoc}
built-up reference Poisson-like or
RMT-like distributions, against the hypothesis that both distributions are
different. In Table \ref{tab:KS} we summarize our results.
We observe that all the spectra emerging from
the three theoretical models are incompatible with RMT, whereas the
experimental one seems to be closer to the Wigner surmise than to the
Poisson distribution. 
Set L2 provides a result that seems to be incompatible with both Poisson
and RMT behaviors. From panel (d) in Fig. \ref{fig:pds_exp} it stems that
the probability to observe spacings closer to zero are higher 
than for a Poisson distribution. Hence, the deviation from 
a Wigner surmise is even larger than from a Poisson distribution.

These results, together with those shown in
Fig. \ref{fig:pds_exp}, point out that the experimental spectrum is
more correlated than the three theoretical ones. Following the work of
Bohigas and Pato \cite{Bohigas04}, we can say that this is
incompatible with the usual statement 
that there are missing resonances in the
experimental spectrum which are included in the theoretical models.
Moreover, the Hamiltonians used in such models do not describe
the statistical properties of the experimental spectrum;
they rather correspond to an integrable system whilst the
experimental spectrum is close to a chaotic system.
Hence, quark models, as they are presently built,
may not be suitable to reproduce the 
low-lying baryon spectrum, and, therefore, 
to predict the existence of missing resonances.

On the other hand, it is important to notice that quark models assume triality
(three-quark states), while QCD allows non-three-quark states
such as pentaquarks and hybrid baryons (excited glue).
Therefore, according to QCD degrees of freedom,
the spectrum which is obtained within quark models cannot be complete.
Consequently, the NNSD of any quark model 
(whatever the interacting Hamiltonian is)
may be deflected to a Poisson
distribution due to its inherent uncompleteness.
The importance and quantification of such effect
depends on the amount of missing 
non-three-quark states present in the actual spectrum and
remains as an open question.
However, we cannot blame 
the non-Wigner character obtained for the quark models studied
in this Letter on this effect:
the loss of correlation cannot be due to missing non-three-quark states
because theoretical models predict more levels than what is observed.

Regardless we obtain that present quark models are not able to predict
missing resonances in the experimental spectrum, our results are
compatible with the existence of some missing states. In fact, the
shape of the NNSD and the Kolmogorov-Smirnov test for a RMT-like
spectrum with a 20\% of random missing levels are alike to those of
set EXP. 

The analysis presented in this Letter
should be extended to baryon spectra
provided by LQCD and EQCDiM as soon as
complete calculations become available. 
In this way we can use the universal properties of fluctuations 
to examine the results given by these models or 
to use the spectra provided by LQCD and EQCDiM 
to test the universality of fluctuations.

\begin{acknowledgments}
The authors thank Dr. T. van Cauteren, Dr. F.J. Llanes-Estrada,
Dr. R.A. Molina, Prof. E. Moya de Guerra,
Dr. J. Retamosa, and Dr. J.M. Ud\'{\i}as for valuable comments.
A.R. is supported by the Spanish program ``Juan de la Cierva''.  
This work has been partly supported under contracts of 
Ministerio de Educaci\'on y Ciencia (Spain)
FTN2003-08337-C04-04 and FIS2005-00640.
\end{acknowledgments}


\begin{thebibliography}{99}
\bibitem{nota1} In Ref. \cite{Molina06} the effects of removing
levels in a non-strictly random way have been studied. It is shown
that if the probability to find a level depends linearly on the
energy, the correlations are roughly lost as predicted
by Bohigas and Pato \cite{Bohigas04}.
\bibitem{nota3} This procedure assumes an energy independent behavior of the
smooth part of the density $\overline{\rho}(E)=1/\left< S \right>$.
The validity of this hypothesis was confirmed
by Pascalutsa in \cite{Pascalutsa03}.
\bibitem{nota2} In doing so, we assume that every sequence $\{ s_i
\}_X$ coming from the same spectra has the same NNSD. This is a
realistic assumption, since we want to calculate the statistical
properties of each spectrum as a whole.
\end{thebibliography}
\end{document}